\documentclass[doublecol]{SOM_based_on_epl2_noheader}

\usepackage{amsmath}
\usepackage{amssymb}
\usepackage{graphicx}
\usepackage{bm}
\usepackage{units}
\usepackage{subfigure}

\title{Diffraction microstrain in nanocrystalline solids under load - heterogeneous medium approach}
\shorttitle{Diffraction microstrain under load}

\author{J\"{u}rgen Markmann\inst{1,2} \and Dmitriy Bachurin\inst{3,4} \and Lihua Shao\inst{2} \and Peter Gumbsch\inst{3,5} \and J\"{o}rg Weissm\"{u}ller\inst{1,2}}
\shortauthor{J\"{u}rgen Markmann \etal}

\institute{
  \inst{1} Universit\"{a}t des Saarlandes, FR7.3 Technische Physik, Saarbr\"{u}cken, Germany\\
  \inst{2} Karlsruher Institut f\"{u}r Technologie, Institut f\"{u}r Nanotechnologie,
           Karlsruhe, Germany\\
  \inst{3} Karlsruher Institut f\"{u}r Technologie, Institut f\"{u}r Zuverl\"{a}ssigkeit von Bauteilen und Systemen,
           Karlsruhe, Germany\\
  \inst{4} Russian Academy of Science, Institute for Metals Superplasticity Problems, Ufa, Russia \\
  \inst{5} Fraunhofer-Institut f\"{u}r Werkstoffmechanik, Freiburg, Germany
}

\abstract{
This is an account of the computation of X-ray microstrain in a polycrystal with anisotropic elasticity under uniaxial external load. The results have been published in the article "Microstrain in nanocrystalline solids under load by virtual diffraction", at \emph{Europhysics Letters} {\bf 89}, 66002 (2010). The present information was submitted to Europhysics Letters as part of the manuscript package, and was available to the reviewers who recommended the paper for publication.}

\begin{document}

\maketitle

\section{Introduction}

This text accompanies a recent publication \cite{EPL} in which the present authors propose a method for computing virtual x-ray diffractograms for nanocystalline materials under uniaxial load based on molecular dynamics simulation data. While the increase in diffraction microstrain during deformation is generally taken as evidence for the generation of lattice dislocations, the virtual diffraction data in Ref. \cite{EPL} show extra microstrain even at small load, before the onset of lattice dislocation activity. Reference \cite{EPL} argues that the microstrain data have a natural explanation in the elastic response of a heterogeneous medium to uniaxial external load. The results imply that the conclusions of previous experimental in-situ diffraction data for nanocystalline metals may deserve a critical examination. 

Here, we account for the computation of X-ray microstrain in a polycrystal with anisotropic elasticity under uniaxial external load from the perspective of the mechanics of heterogeneous elastic media.

\section{General approach to microstrain}

We consider a space-filling array of grains with random crystallographic orientation, in a way that the polycrystalline material is macroscopically isotropic, and we restrict attention to purely elastic deformation. We allow for anisotropic elastic response of the crystal lattice of the individual grains, consistent with cubic symmetry. In a diffraction experiment the heterogeneous strain in the polycrystal will lead to broadening of the Bragg reflections which is measured by the diffraction microstrain, as described below.

Our analysis uses crystal coordinates. Consider diffraction at a Bragg reflection with Miller indices $hkl$, and let the vectors $\vec{u}_{hkl}$, $\vec{v}_{hkl}$ and $\vec{w}_{hkl}$ form an orthonormal basis set with ${\vec u}_{hkl} = (h,k,l)/\sqrt{h^2+k^2+l^2}$ the unit vector along the zone axis that is aligned with the scattering vector, ${\vec q}$. For instance, for the (422)-reflection one may take ${\vec u} = (2,1,1)/\sqrt{6}$, ${\vec v} = (0,-1,1)/\sqrt{2}$ and ${\vec w} = (-1,1,1)/\sqrt{3}$. Since the diffraction experiment has ${\vec q}$ normal to the load axis, a general expression for the unit vector ${\vec n}$ along the load axis is
\begin{equation}
\label{load_axis}
{\vec n}_{hkl}(\varphi)
=
{\vec v}_{hkl} \cos \varphi + {\vec w}_{hkl} \sin \varphi
\end{equation}
with $\varphi$ an orientation variable. In other words, at any given Bragg reflection $hkl$ the crystals which satisfy the Bragg condition are responding to an external load that can take on different directions as seen within a crystal coordinate system. When $\varphi$ varies between $0$ and $2\pi$, then the vector ${\vec n}_{hkl}(\varphi)$ covers the entire domain of relevant orientations of the external load axis. The uniaxial stress due to the respective load is given by the tensor
\begin{equation}
\label{uniaxial_load}
{\bf T}_{hkl}(\varphi)
=
\sigma \; {\vec n}_{hkl}(\varphi) \otimes {\vec n}_{hkl}(\varphi) \,.
\end{equation}
Here, $\sigma$ denotes the magnitude of the stress.

The interference function of the polycrystalline aggregate is the superposition of Bragg reflections of crystals with a variety of load directions in crystal coordinates, as discussed above. The Bragg reflection position of any given crystallite is determined by the lattice parameter, measured in the direction of ${\vec q}$. Since the strain -- in other words, the relative change of the lattice parameter as compared to a stress-free reference configuration -- depends on the load direction, there will generally be a distribution of lattice parameters and, hence, a distribution of reflection positions. It is this fact that links the reflection broadening to the load via the anisotropic elastic response of the crystal lattice. 

Specifically, the reflection broadening is measured by the microstrain, $e_{hkl}$, which depends on the projection, $\epsilon_{hkl}(\varphi)$, of the strain $\bf E$ onto ${\vec u}_{hkl}$ via
\begin{equation}
\label{variance_strain}
e_{hkl}^2
=
\frac 1 {2 \pi}
\int_0^{2 \pi}
\left(
\epsilon_{hkl}(\varphi)
-
\langle \epsilon_{hkl} \rangle
\right)^2
\rm d \varphi
\end{equation}
In the following we compute $e_{hkl}^2$ using first the Reuss and then the Kr\"oner approximation.

\section{Reuss approximation}

As a first step towards analysing the strain, consider the Reuss approximation, where the stress is assumed uniform throughout the polycrystal. The local stress in any grain is then given by equation (\ref{uniaxial_load}), and the strain is
\begin{equation}
\label{local_strain}
{\bf E}_{hkl}^{\rm R}(\varphi)
=
{\mathbf S} : {\bf T}_{hkl}(\varphi)
\end{equation}
with ${\bf S}$ the compliance tensor. The superscript refers to the Reuss aproximation. The strain projected on the
scattering vector is here
\begin{equation}
\label{projected_strain_Reuss}
\epsilon_{hkl}(\varphi)
=
{\vec u}_{hkl} \cdot {\bf E}_{hkl}^{\rm R}(\varphi) \cdot {\vec u}_{hkl} \,.
\end{equation}

By using equations (\ref{local_strain}) and (\ref{projected_strain_Reuss}) along with tabulated values for the single-crystal compliance coefficients (represented, for instance, by $C_{11}$, $C_{12}$, $C_{44}$ for a cubic crystal lattice), equations (\ref{variance_strain}) for the square of the microstrain value $e_{hkl}$ is readily evaluated at each Bragg reflection.

\section{Kr\"{o}ner approximation}

To evaluate the microstrain in the Kr\"{o}ner approximation \cite{Kroener1958,Hutchinson1970} we take the grain of orientation $h,k,l, \varphi$ as an inclusion that is embedded in an elastically isotropic continuum with the self-consistently averaged (Kr\"{o}ner-) elastic constants, represented here by ${\bf \tilde S}$. The entries of the tensor ${\bf \tilde S}$ can be computed by means of equations (21) and (22) in reference \cite{Kroener1958}. Sufficiently far from the inclusion, the continuum is then strained by
\begin{equation}
\label{matrix_strain}
{\bf {\tilde E}}_{hkl}(\varphi)
=
{\bf \tilde S} : {\bf T}_{hkl}(\varphi) \, .
\end{equation}
The strain within the grain is uniform, and it is given by (cf. equation (14) in reference \cite{Kroener1958})
\begin{equation}
\label{misfit_strain}
{\bf E}_{hkl}^{\rm K}(\varphi)
=
\left( {\bf \tilde S} + {\bf t}_{hkl}(\varphi) \right): {\bf T}_{hkl}(\varphi) \, .
\end{equation}
The quantity ${\bf t}_{hkl}(\varphi)$ describes an excess compliance due to the interaction between grain and matrix. It can be computed from the single-crystal elastic constants by means of equations (18) and (19) in reference \cite{Kroener1958}.

Analogously to equation (\ref{projected_strain_Reuss}), the strain projected on the scattering vector is here
\begin{equation}
\label{projected_strain_Kroener}
\epsilon_{hkl}(\varphi)
=
{\vec u}_{hkl} \cdot {\bf E}_{hkl}^{\rm K}(\varphi) \cdot {\vec u}_{hkl} \,.
\end{equation}

By using equations (\ref{local_strain})-(\ref{projected_strain_Kroener}) along with the single-crystal compliance coefficients, we can again evaluate equation (\ref{variance_strain}) for the square of the microstrain value $e_{hkl}$ at each Bragg reflection.

\section{Results}

The microstrain values $e_{hkl}$ computed from our model are linear in the stress magnitude, $\sigma$. We can therefore specify a stress-independent microstrain parameter, $b$, defined so that
\begin{equation}
e_{hkl} = b_{hkl} \sigma / Y \,.
\end{equation}
Here $Y$ is taken as the macroscopic Young modulus of the respective model, Reuss or Kr\"oner.

Table \ref{t1} shows the values obtained with the Reuss and Kr\"{o}ner models. Values for the single crystal stiffness were those of the Sydow potential for Pd \cite{vonSydow1999}, namely $C_{11} = \unit[292.7]{GPa}$, $C_{12} = \unit[188.9]{GPa}$, $C_{44} = \unit[125.0]{GPa}$. The corresponding Kr\"oner elastic constants were obtained as $Y = \unit[235]{GPa}$, $G = \unit[88.7]{GPa}$, $\nu = 0.325$ where $Y, G, \nu$ denote the Young modulus, the shear modulus, and the Poisson ratio, respectively. .

\begin{table}
\caption{\label{t1}
    The dimensionless microstrain coupling coefficients $b$ for Pd (Sydow potential) in different crystallographic orientations $(hkl)$. Values derived from Reuss ($b^{\rm R}$) and Kr\"oner ($b^{\rm K}$) models are distinguished by superscripts. The bottom line shows the root-mean square of the coefficients.
   }
\begin{center}
\begin{tabular}
{c|c|c}
\hline\hline
$hkl$ &  $b^{\rm R}$ &  $b^{\rm K} $
\\\hline
(100) & 0 & 0
\\
(110) & 0.213 & 0.106
\\
(111) & 0 & 0
\\
(210) & 0.137 & 0.068
\\
(211) & 0.071 & 0.035
\\
(221) & 0.126 & 0.063
\\
(310) & 0.077 & 0.038
\\
(311) & 0.056 & 0.038
\\
(321) & 0.130 & 0.065
\\
(331) & 0.170 & 0.085
\\\hline
RMS & 0.118 & 0.059
\\\hline\hline
\end{tabular}
\end{center}
\end{table}

\end{document}